Title:  **Molecular Constraints on Synaptic Tagging and Maintenance of Long-Term Potentiation: A Predictive Model**

Authors:   Paul Smolen, Douglas A. Baxter, and John H. Byrne

Laboratory of Origin:

Department of Neurobiology and Anatomy

W. M. Keck Center for the Neurobiology of Learning and Memory

The University of Texas Medical School at Houston

Houston, Texas 77030

United States of America

Correspondence Address:

Paul Smolen

Department of Neurobiology and Anatomy

The University of Texas-Houston Medical School

Houston, TX 77030

E-mail: Paul.D.Smolen@uth.tmc.edu

Voice: (832) 859-3826

## ABSTRACT

Protein synthesis-dependent, late long-term potentiation (LTP) and depression (LTD) at glutamatergic hippocampal synapses are well characterized examples of long-term synaptic plasticity. Persistent increased activity of protein kinase M $\zeta$ (PKM$\zeta$) is thought essential for maintaining LTP. Additional spatial and temporal features that govern LTP and LTD induction are embodied in the synaptic tagging and capture (STC) and cross capture hypotheses. Only synapses that have been "tagged" by a stimulus sufficient for LTP and learning can "capture" PKM$\zeta$. A model was developed to simulate the dynamics of key molecules required for LTP and LTD. The model concisely represents relationships between tagging, capture, LTD, and LTP maintenance. The model successfully simulated LTP maintained by persistent synaptic PKM$\zeta$, STC, LTD, and cross capture, and makes testable predictions concerning the dynamics of PKM$\zeta$. The maintenance of LTP, and consequently of at least some forms of long-term memory, is predicted to require continual positive feedback in which PKM$\zeta$ enhances its own synthesis only at potentiated synapses. This feedback underlies bistability in the activity of PKM$\zeta$. Second, cross capture requires the induction of LTD to induce dendritic PKM$\zeta$ synthesis, although this may require tagging of a nearby synapse for LTP. The model also simulates the effects of PKM$\zeta$ inhibition, and makes additional predictions for the dynamics of CaM kinases. Experiments testing the above predictions would significantly advance the understanding of memory maintenance.

## AUTHOR SUMMARY

A fundamental problem in neurobiology is to understand how memories are maintained for up to years. Long-term potentiation (LTP), an enduring increase in the strength of specific connections (synapses) between neurons, is thought to comprise, at least in part, the substrate of learning and memory. What processes transduce brief stimuli into persistent LTP? Persistent increased activity of an enzyme denoted protein kinase M $\zeta$ (PKM$\zeta$) is thought essential for maintaining LTP. Only synapses that have been "tagged" by a stimulus, such as stimuli needed for LTP and learning, can "capture" PKM$\zeta$. We developed a model simulating dynamics of key molecules required for LTP and its opposite, long-term depression (LTD). The model concisely represents relationships between tagging, capture, LTD, and LTP maintenance. It makes testable predictions concerning the dynamics of PKM$\zeta$. The maintenance of LTP and memory is predicted to require positive feedback in which PKM$\zeta$ enhances its own synthesis at potentiated synapses. Without synaptic capture of PKM$\zeta$, no positive feedback would occur. LTD induction is also predicted to increase PKM$\zeta$ synthesis. The model also makes predictions about regulation of PKM$\zeta$ synthesis. Experiments testing the above predictions would advance the understanding of memory maintenance.

**INTRODUCTION**

Protein synthesis-dependent forms of LTP and LTD (late LTP/D, henceforth abbreviated LTP and LTD) are the subject of intense study because they represent cellular mechanisms of long-term memory. Some key mechanisms underlying the induction and maintenance of LTP and LTD are emerging. These include compartmentalization, within stimulated dendritic spines, of $Ca^{2+}$ signals and of kinase activation [1,2], and synapse specificity of induction mediated by synaptic tagging and capture (STC). In STC [3,4], one synapse (S1) receives either a weak high-frequency tetanus (WTET) or a weak low-frequency stimulus (WLFS). Neither WTET nor WLFS induce LTP or LTD. However, such stimuli "tag" the activated synapse for subsequent plasticity. Consequently, if activity in S1 closely precedes or follows a strong tetanus (STET) or strong low-frequency stimulus (SLFS) at a second synapse (S2), long-term changes occur at S1.

To establish LTP/D, dendritic translation of plasticity-related proteins (PRPs) follows the strong S2 stimulus. The tag allows capture of PRPs at S1. The direction of plasticity at S1 is determined by the type of tag, established by WTET (LTP) or WLFS (LTD). Therefore, PRPs generated by either STET or SLFS are able to support either LTP or LTD, and the tag at S1 determines whether LTP or LTD occurs [3-7]. In a cross capture (or cross tagging) protocol, LFS at one synapse is paired with tetanus at the other synapse [3,7]. If WLFS at S1 tags S1 for LTD, then LTD occurs subsequent to STET at S2. Conversely, WTET at S1 yields LTP when paired with SLFS at S2. The autonomously active isoform of atypical protein kinase C ζ, termed protein kinase M ζ (PKMζ), is a PRP. PKMζ activity is necessary for induction and maintenance of at least some forms of LTP [8-10] and for induction of PKMζ synthesis [11]. For brevity, we henceforth denote PKMζ as simply "PKM".

How are the above processes integrated into a synapse that can express different forms of plasticity? To help in understanding the integrated function, we developed a computational model that describes some of the postsynaptic molecular cascades at hippocampal CA3-CA1 synapses. In order to simulate induction and maintenance of LTP and LTD, STC, and cross capture, several key assumptions, each consistent with empirical data, were necessary. In the model, kinases are differentially regulated in a dendritic compartment *vs.* a synaptic compartment, with the latter corresponding to a stimulated spine. The synthesis of PKM is likewise differentially regulated, in order to simulate cross capture in which WTET induces LTP when paired with SLFS. Either STET or SLFS stimulates synthesis of PKM in the dendritic compartment. This synthesis is regulated by a CaM kinase, possibly CaM kinase II. PKM is only captured by the synaptic compartment if an LTP tag is set. Maintenance of LTP is mediated by bistability in PKM activity restricted to the synaptic compartment. Persistent PKM activation is sustained by positive feedback in which synaptic PKM enhances its synthesis. These model assumptions result in testable empirical predictions for the dynamics of PKM, CaM kinases, and synaptic tags.

**RESULTS**

**Simulation of STET–induced LTP and SLFS–induced LTD**

The model incorporates postsynaptic and dendritic roles for the MAP kinase isoform(s) termed extracellular-regulated kinase (ERK), CaM kinases, and a phosphatase necessary for LTD. Synthesis rates of PRPs are described by saturable functions of the concentrations of phosphorylated kinase targets. These targets could represent translation factors. Setting of synaptic tags is also described by phosphorylation/dephosphorylation of targets. LTP or LTD corresponds to increases or decreases in a synaptic weight W. Fig. 1 schematizes the signaling cascades that lead from synaptic and dendritic stimuli to increases or decreases in W.

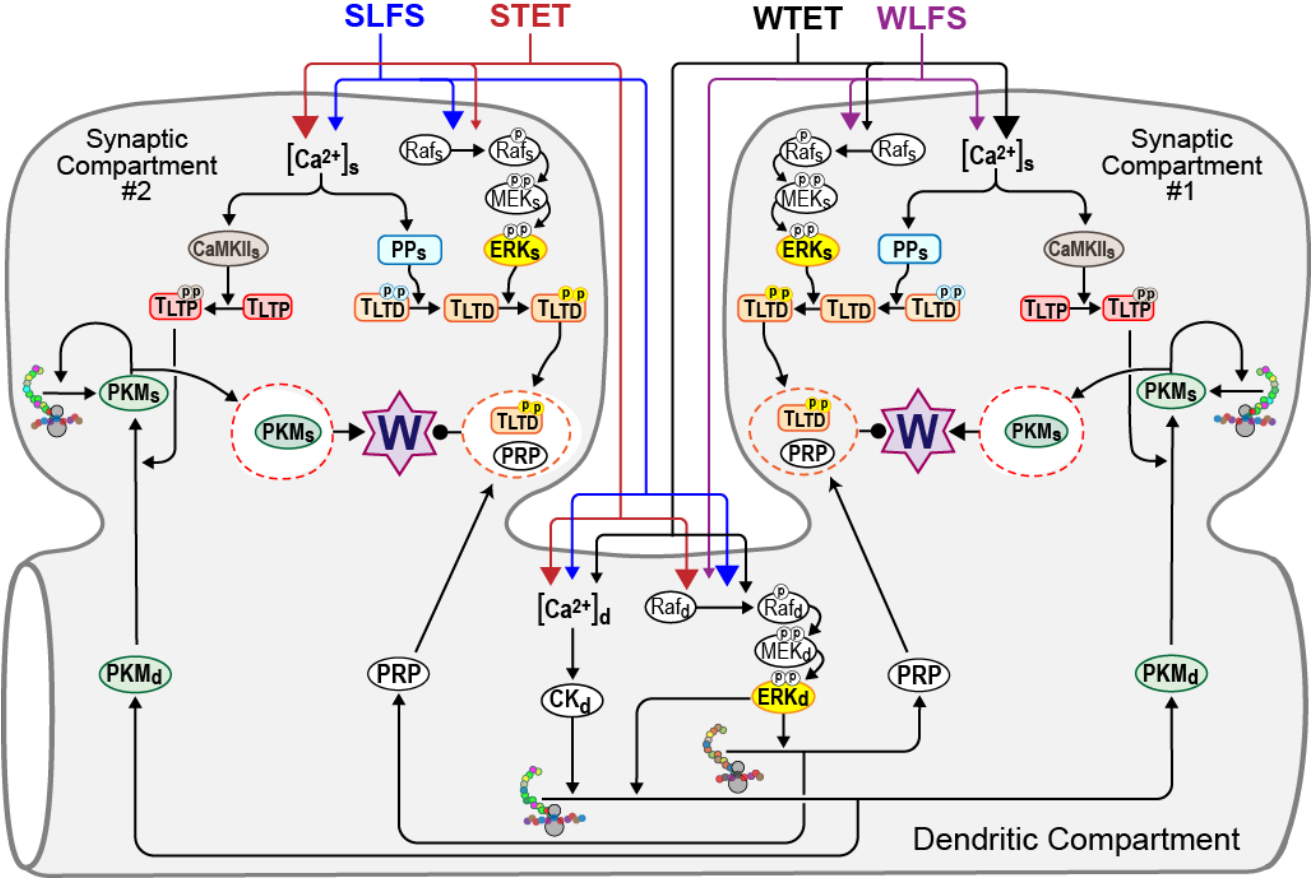

FIGURE 1  Schematic representations of two synaptic spines (Synaptic Compartments #1 and #2) and an adjacent dendritic shaft (Dendritic Compartment). The biochemical cascades in Synaptic Compartments #1 and #2 are identical.  The two compartments represent, respectively, the response of the model to weak *vs.* strong stimuli.  The stimulus protocols are:  *i*) Strong low-frequency stimulus (SLFS);  *ii*) Strong tetanic stimulus (STET);  *iii*) Weak tetanic stimulus (WTET); and  *iv*) Weak low-frequency stimulus (WLFS). Stimuli increase levels of Ca$^{2+}$ ([Ca$^{2+}$]$_s$ and [Ca$^{2+}$]$_d$) and activate synaptic and dendritic ERK cascades (denoted, respectively, Raf$_s$/MEK$_s$/ERK$_s$ and Raf$_d$/MEK$_d$/ERK$_d$).  Sizes of arrowheads reflect the impact that each protocol has on Ca$^{2+}$ levels and on ERK activation.  The model includes three novel features.  First, translation of dendritic PKM, PKM$_d$, requires activity of ERK$_d$ and an unidentified, Ca$^{2+}$-dependent kinase (CK$_d$).  Second, PKM$_d$ can only be captured by a synaptic compartment if that synapse is tagged for LTP.  Third, bistability and persistent activation of synaptic PKM, PKM$_s$, occurs only in the synaptic compartment.

We simulated LTP induction by STET (three 1-s tetani) (Fig. 2, see Methods for model equations and parameters).  CaMKII$_s$ remains active for ~5 min (Fig. 2A).  Raf activation leads to activation of synaptic and dendritic ERK, ERK$_s$ and ERK$_d$.  The activation lasts ~90 min (Figs. 2A-B).  Substantial amounts of PKM and PRP are synthesized due to ERK$_d$ and CK$_d$ activation. The LTP tag, T$_{LTP}$, peaks in ~1 min after the third stimulus, close to its upper bound of 1.  LTP nears completion in ~2 h (Fig. 2C, time course of W).  Empirically, induction of LTP with BDNF/forskolin (bypassing E-LTP) requires 1-2 h [12].  The upper state of W is stable.  Five h post-tetanus, W is close to this state and is elevated 170% above baseline.  This magnitude is similar to EPSP increases observed after multiple tetani [13].

In cross capture experiments that pair a WTET with an SLFS, LTP instead of LTD is observed at the tetanized synapse.  Therefore, a tetanus does not set an LTD tag.  In the model, in order to prevent T$_{LTD}$ from increasing substantially in response to tetani while also simulating LTD and STC, it was necessary to assume a tetanus activates synaptic Raf, and consequently synaptic ERK, much less than dendritic Raf and ERK (compare the $ERK_s$ and $ERK_d$ time courses in Figs. 2A-B) (see Methods for stimulus parameters). Consequently, there is little phosphorylation of the $ERK_s$ $T_{LTD}$ site. $T_{LTD}$ remains low (note scale factor of 100 in Fig. 2A). $ERK_d$, in conjunction with $CK_d$, drives synthesis of $PKM_d$.

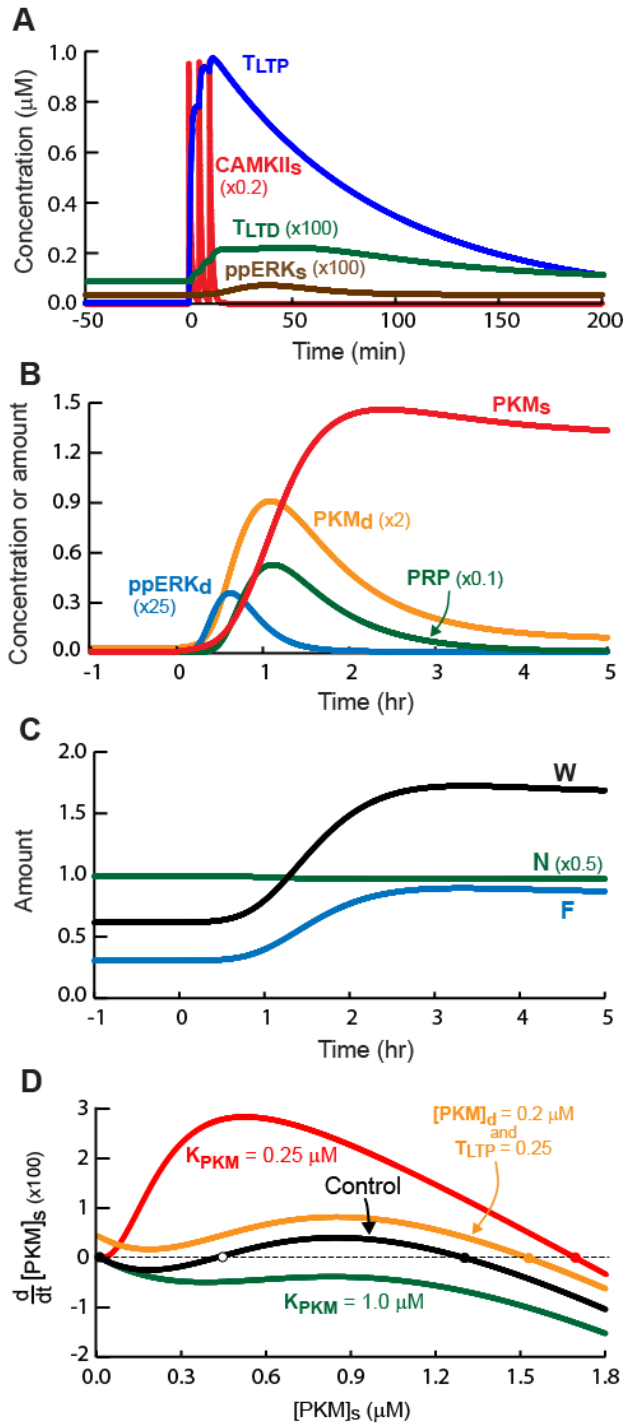

FIGURE 2 Simulated induction of LTP by STET. **A,** Time courses of CaMKII$_s$ activity, ERK$_s$ activity, and the LTP and LTD tags. ERK$_s$ activity is the concentration of the doubly phosphorylated, active form of ERK$_s$, ppERK$_s$. To help compare the dynamics of the variables, which differ greatly in amplitude, the vertical scales for ERK$_s$ and T$_{LTD}$ are multiplied by 100, and the scale for CaMKII$_s$ is multiplied by 0.2. **B,** Time courses of ERK$_d$ activity, PRP, PKM$_s$, and PKM$_d$. ERK$_d$ is vertically scaled by a factor of 25, PRP by 0.1, and PKM$_d$ by 2. **C,** Time courses of N, F, and W. **D,** Bistable switch of PKM$_s$. The derivative is vertically scaled by 100. Stable steady states are indicated by filled circles, an unstable steady state by an open circle. The control plot (black curve) is with the standard parameter values, and has two stable steady states, at PKM$_s$ = 0.0096 µM and at PKM$_s$ = 1.30 µM. For K$_{PKM}$ ≥ 0.87 µM, only the lower stable steady state is present (green curve). For K$_{PKM}$ ≤ 0.25 µM, only the upper stable state is present (red curve). The lower state is also eliminated if PKM$_d$ and T$_{LTD}$ are increased (orange curve).

Figure 2B illustrates PKM dynamics. $PKM_d$ starts at a low basal level and $PKM_s$ is even lower due to the absence of $T_{LTP}$ and thus the absence of translocation of PKM. After STET, $PKM_d$ peaks in ~1 h. This time is consistent with the empirical time course of PKM levels in hippocampal slice [11,14]. However, the initial rise in $PKM_d$ takes ~30 min to develop, whereas the empirical PKM increase is significant somewhat earlier, at 10 min [14]. Therefore, the model does not completely represent the early dynamics of PKM synthesis.

As a consequence of PKM translocation, $PKM_s$ increases. When this increase is comparable to the Hill constant $K_{pkm}$ in Eq. 29, positive feedback is initiated in which $PKM_s$ activates its own synthesis. $PKM_s$ converges to the upper state of a bistable switch. $PKM_s$ takes ~2 h to reach peak. In the dendrite, positive feedback does not operate, so $PKM_d$ declines. The rate constant for PKM degradation, $k_{d\_PKM}$, is 0.02 min⁻¹, corresponding to a half-life of 35 min. However, because synthesis of $PKM_d$ does not terminate abruptly in Fig. 2, the decline of $PKM_d$ takes place over a few h. The increase in $PKM_s$ drives a sustained increase in the variable F (Fig. 2C), which represents the amount of available, phosphorylated AMPA receptors that are functionally incorporated into postsynaptic sites. The variable N, representing the number of receptors available for incorporation, does not change substantially (Fig. 2C), because N remains near its basal value unless concurrent elevation of the LTD tag and PRP occurs. The synaptic weight W is given as the product of F and N. W and F transit from a lower to an upper state (Fig. 2C), and remain elevated.

Figure 2D illustrates the effect on bistability of varying the strength of positive feedback. The switch is visualized by plotting $PKM_s$ on the $x$-axis and its derivative on the $y$-axis. To represent persistent PKM activity after the LTP tag decays, the plot is with $T_{LTP} = 0$, so there is no influx of PKM from the dendrite. The black curve is with the Hill constant of feedback, $K_{pkm}$, at its standard value, 0.75 μM. Three steady states are seen where the derivative of $PKM_s$ is zero. The left and right states (filled circles) are stable to small perturbations of $PKM_s$, the middle state (open circle) is unstable. At the stable states the curve has negative slope, so that a small increase (decrease) of $PKM_s$ will yield a negative (positive) derivative, relaxing $PKM_s$ back to the steady state. An increase in $K_{pkm}$ represents a decrease in the feedback strength, because more $PKM_s$ is required to activate its own synthesis. For $K_{pkm} \geq 0.87$ only the lower stable state is present, with the derivative zero at a single low value of $PKM_s$. In contrast, as $K_{pkm}$ decreases, the feedback strength increases until only the upper stable state is present. Bistability can also be eliminated by influx of PKM from the dendrite. With $PKM_d$, set to 0.2 μM, then for $T_{LTP} > 0.23$, influx of PKM eliminates the lower steady state of $PKM_s$. During induction of LTP, the coincident increase of $PKM_d$ and $T_{LTP}$ similarly drives PKM influx and eliminates the lower state.

*In vitro* (and *in vivo*), application of a PKM inhibitor, ZIP, several h after LTP induction (or behavioral training) irreversibly abolishes LTP (as well as several forms of long-term memory) [8,15,16,17]. A second PKM inhibitor, chelerythrine, similarly eliminates several forms of LTM [17]. To simulate ZIP's effect on LTP, the effect of a strong (80%) PKM inhibition was simulated by multiplying the activity of $PKM_s$ by 0.2. That is, in the first terms on the RHS of Eqs. 29 and 31, $PKM_s$ was multiplied by 0.2 and its square by 0.04. Given that the physiological interactions of PKM with its substrate(s) have not been characterized, this simple method of simulating inhibition appears reasonable. Figure 3 illustrates that 1 h of inhibition, beginning 5 h after LTP induction, returns $PKM_s$, F, and W to their stable lower states. The empirical loss of LTP occurs within ~1 h of the start of PKM inhibition. Therefore, in the model, the time constant of F ($\tau_F$) was chosen to be relatively rapid (30 min) so that the return of F and W to their lower states nears completion in 1 h. A weaker simulated inhibition of PKM (30% for 1 h) generates only a temporary dip in $PKM_s$ and W.

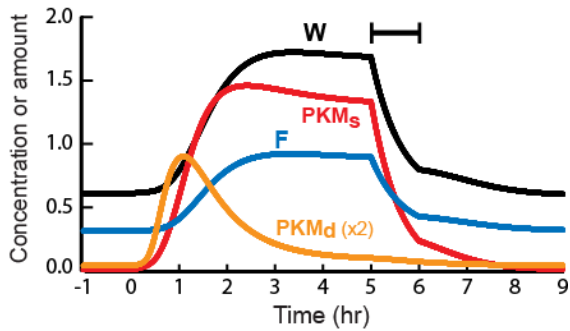

FIGURE 3  Simulated block of LTP maintenance by temporary inhibition of PKM.

Simulated inhibition of CaMKII during STET also blocks LTP (CaMKII activity was reduced by 85% for 10 min, starting at the beginning of STET). Also, simulated inhibition of MEK activation during and immediately after STET blocks LTP. This simulation corresponds to the action of the common MEK inhibitors U0126 and PD98059, and was implemented by an 80% reduction in the activation rate constant $k_{p\_MEK}$ for 10 min starting at the beginning of STET. In the model, the maintenance of established LTP only depends on elevated $PKM_s$ (Eqs. 31-32), and is not affected by inhibition of other kinases.

SLFS, which induces LTD, was simulated as a 15-min elevation of Raf activation and $Ca^{2+}$ (details in Methods). $CaMKII_s$ activates only slightly (Fig. 4A), because synaptic $Ca^{2+}$ is lower than with STET. This slight $CaMKII_s$ activation yields a small elevation of $T_{LTP}$ at S1. $CK_d$ activates strongly due to its lower $Ca^{2+}$ threshold. $CK_d$, together with $ERK_d$, drives $PKM_d$ synthesis (Fig. 4B). An alternate model would postulate PKM synthesis is only upregulated when SLFS is paired with a WTET in a cross capture protocol (see Discussion).

In Fig. 4, $T_{LTP}$ is not significantly elevated, so little capture of PKM into the synaptic compartment occurs (Fig. 4B shows only a small transient elevation of $PKM_s$). As a result, PKM fails to drive incorporation of synaptic AMPARs. F is only slightly elevated (Fig. 4C). PRP is synthesized due to activation of $ERK_d$. With standard parameter values, synthesis of PRP and $PKM_d$ due to the relatively long SLFS stimulus exceeds that due to the strong but brief tetani in STET. The prolonged SLFS also activates $ERK_s$, increasing $T_{LTD}$, which peaks at 0.16, a ~40-fold increase over basal $T_{LTD}$. The variable N, representing the number of AMPARs that are available to be incorporated at the synapse, is decreased (Fig. 4C) due to concurrent elevation of PRP and $T_{LTD}$ (Eq. 30). Assessed 3 h after stimulus, LTD of 51% was simulated (W decreases by 51% from its basal value) (Fig. 4C). Empirically, LTD of ~50% is near saturation, and conversely LTP is rarely observed to exceed 200% [13].

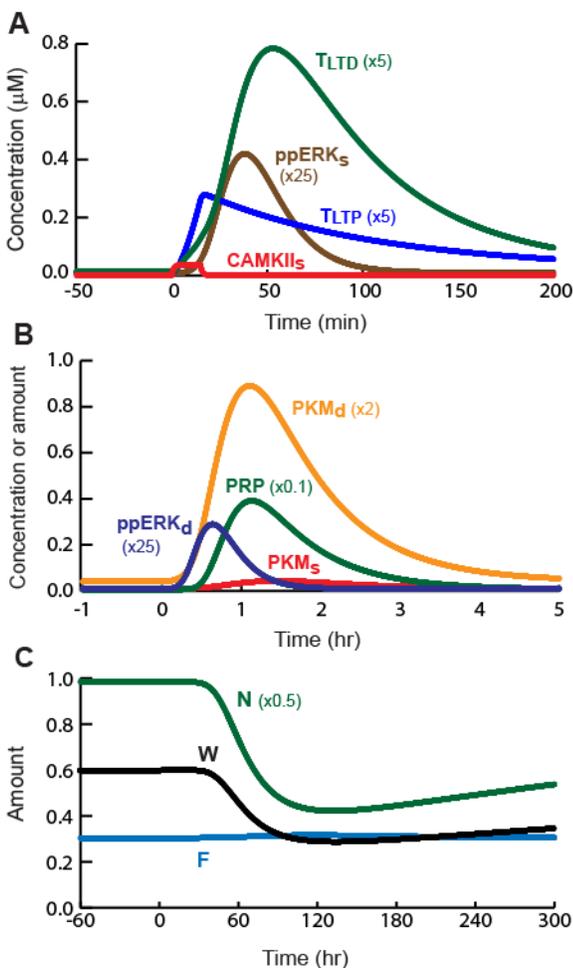

FIGURE 4 Simulated induction of LTD by the SLFS protocol. The traces of $ppERK_s$, $T_{LTD}$, $ppERK_d$, and $PKM_s$ are vertically scaled as indicated.

Empirically, WTET and WLFS set LTP and LTD tags respectively. In the model, WTET and WLFS activate, respectively, $CaMKII_s$ and $ERK_s$. In Fig. 2A, the time course of $T_{LTP}$ has three peaks due to the tetani. The first tetanus, equivalent alone to WTET, largely saturates $T_{LTP}$, which peaks at 0.77, 3 min post-tetanus. Synaptic WLFS stimulus parameters are the same as for SLFS, so $T_{LTD}$ responds as in Fig. 4A.

Empirically, although WTET does not generate LTP, a single theta-burst does [12]. This stimulus consists of 10-12 bursts of four 100 Hz pulses, 200 msec apart (total duration ~ 2.2 s). With the model, to simulate LTP induction by this single stimulus, the dendritic Raf activation profile was given the same shape as for a tetanus. However, the LTP state transition only occurred if this profile was scaled up substantially, by increasing the maximal amplitude of $k_{p\_Raf,d}$ to 0.08 $min^{-1}$ (The $Ca^{2+}$ elevation had the same amplitude as for WTET but lasted 4 s). A previous model [18] similarly suggested theta-bursts may be particularly effective at activating dendritic ERK.

An important consideration for biochemical models is the sensitivity of dynamics to the values of parameters. This issue was examined by repeating the simulations of Figs. 2 and 4. For each of the 49 parameters, the standard value (Methods) was varied by +20% and also by -20%. For each of these 98 variations, W was assessed 3 h after STET and also 3 h after SLFS. After STET, the mean value of W was 1.63 with a standard deviation (SD) of 0.12. After SLFS, the mean of W was 0.29 with an SD of 0.07. The magnitude of these SDs relative to the means suggests that overall, the model is not unduly sensitive to significant parameter variations.

However, a relatively high sensitivity of LTD to some of the parameters in the ERK cascade was observed. For variations of $k_{p\_MEK}$, $k_{dp\_MEK}$, and $K_{MEK}$, the magnitude of W following SLFS varied by more than 50% from the control simulation of Fig. 4. In an attempt to moderate this sensitivity, we repeated the LTD simulations for these parameters but varied them separately for the dendritic and synaptic ERK cascades. In 6 simulations, these 3 parameters were held constant in the synaptic compartment but varied by +20% and -20% in the dendritic compartment, and vice versa in 6 additional simulations. This procedure reduced sensitivity, but the maximal and minimal values of W were respectively still 42% greater than control and 35% less than control. Thus, future examination of model variants with reduced sensitivity of LTD to ERK cascade parameters appears warranted. For example, the synthesis of PRP might be made dependent on phosphorylations by more than one kinase, rather than dual ERK phosphorylations.

The effect of larger variations in the parameter $V_{sd}$, the ratio of volumes of the synaptic to dendritic compartments, was also considered. $V_{sd}$ affects PKM dynamics (Eqs. 28-29). Therefore, variations in $V_{sd}$ will alter, and potentially eliminate, bistability in $PKM_s$ and in W. However, the values of $PKM_s$ in both stable states and the response to STET were robust to substantial variations in $V_{sd}$. Reducing $V_{sd}$ from its standard value of 0.03 to 0.01 preserved an STET response very similar to the LTP illustrated in Fig. 2. The values of $PKM_s$ in both stable states were preserved to within 5%. The only difference in dynamics was a substantial (~90%) transient overshoot in $PKM_s$, and an overshoot in W, above their upper states. If $V_{sd}$ was instead increased, to 0.1, bistability was again preserved, with similar values of $PKM_s$ in both states. However, because of the larger synaptic volume, STET no longer drove a state transition unless another parameter was also varied to increase PKM influx into the synapse. Increasing the rate constant $k_{s \to d}$ 3-fold (to 0.0075 $min^{-1}$) restored the transition.

**Two-compartment PKM dynamics appear necessary to simulate STC**

In tagging protocols, S1 receives WTET or WLFS and S2 receives STET or SLFS. PRP and $PKM_d$ synthesis is predominantly driven by the S2 stimulus. Only after both stimuli are paired will LTD occur at S1 (if $T_{LTD}$ was set by the S1 stimulus), or LTP occur (if $T_{LTP}$ was set). A "weak before strong" STC experiment [6] was simulated with WTET to S1 followed 20 min later by STET to S2. Figure 5A1 illustrates time courses at S1 of $T_{LTP}$ and W. Strong LTP occurs (a 178% increase in W, 5 h post-stimulus). The time course of $PKM_s$ is also shown. The PKM captured at S1 is mostly generated by the STET at S2. Empirically, STET at S2 does not generate LTP at S1 unless the LTP tag at S1 is set [16], and the elevation of PKM after LTP induction appears punctate, possibly restricted to stimulated spines, rather than uniform throughout a section of dendrite [19]. These data support the representation of PKM dynamics in the model, in which a) PKM is synthesized dendritically in response to strong stimuli, b) the tag allows capture into a synaptic compartment, and c) the biochemical switch that yields persistent PKM activation is restricted to the synaptic compartment.

STC with an LTD protocol was also simulated [4,20] with WLFS to S1 followed, after 5 min, by SLFS to S2. Figure 5B1 illustrates time courses of $T_{LTD}$, $PKM_s$, and W. WLFS does not elevate $Ca^{2+}$ enough to activate $CaMKII_s$ substantially, so $T_{LTP}$ remains low. Thus only a small, transient increase in $PKM_s$ occurs. However, PRP is elevated due to the SLFS at S2, and is captured at S1 due to the elevated $T_{LTD}$. Therefore strong LTD occurs (a 53% decrease in W, 3 h after SLFS).

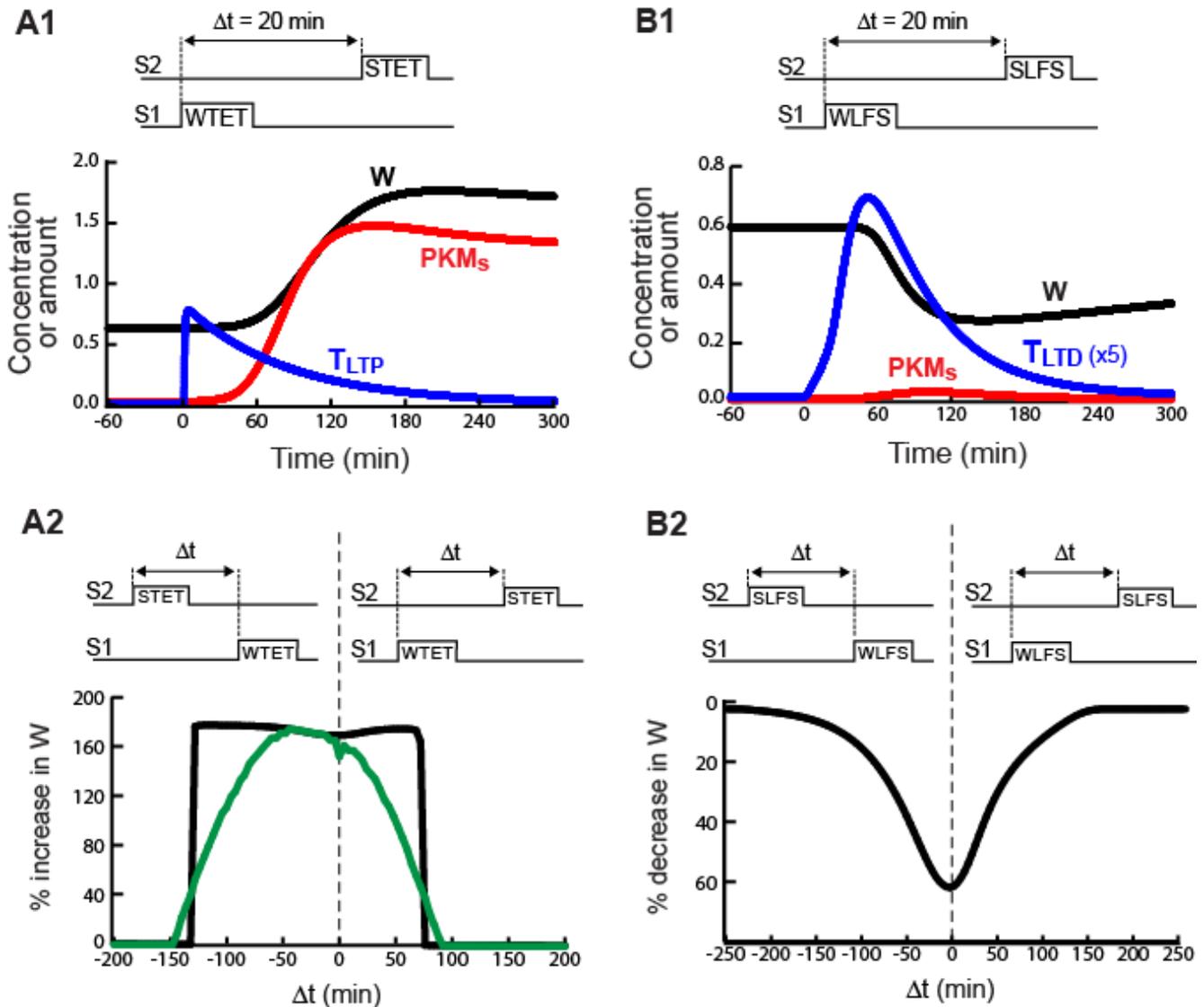

FIGURE 5  Simulation of STC. **A1,** A WTET is delivered to S1 20 min prior to the start of STET delivered to S2. Illustrated synaptic variables are for S1. If WTET is followed by STET, LTP occurs. **A2,** STC for LTP occurs during a limited time window. The *x*-axis represents time between the beginning of the STET and WTET. A negative Δt indicates that STET precedes WTET. The *y*-axis represents the percent change of W at S1, assessed 5 h after the WTET to S1. 5 h sufficed for LTP to approach its steady-state value. Black curve, time window for a single synaptic compartment coupled to a dendritic compartment. Green curve, time window for a population of 40 compartment pairs. Over the 40 pairs, the value of the rate constant $k_{d \to s}$ increased in uniform steps from 25% to 125% of its standard value of 0.0025 min$^{-1}$. LTP peaks when STET precedes WTET by 44 min. **B1,** A WLFS is delivered to S1 20 min prior to SLFS at S2. If WLFS is followed by SLFS, LTD occurs. **B2,** STC for LTD occurs during a limited time window. LTD is assessed 3 h after the second of the paired stimuli.

Figures 5A2 and 5B2 show simulated timing windows for the induction of LTP and LTD. The *x*-axes represent the intervals by which a strong stimulus to S2 precedes (negative intervals) or follows (positive intervals) a weak stimulus to S1. The *y*-axes represent the induced percent change of W at S1. In Fig. 5A2, the black curve is the timing window for LTP with a single synaptic compartment S1. LTP is assessed 5 h after STET to allow W to converge to the stable upper state. LTP occurs if WTET precedes STET by 75 min or less. This maximal interval is somewhat less than reported, in that [6] observed a minor component of LTP remaining at 2 h. LTP also occurs if WTET follows STET by 125 min or less. This maximal interval does not seem to have been empirically investigated.

Because of the bistable switch in $PKM_s$, this LTP window has an abrupt rise and fall. Either LTP is complete ($PKM_s$ and W switch to their upper states) or it does not occur at all ($PKM_s$ and W return to the lower state after a transient). Empirically, however, the LTP window exhibits a sloped rise and fall [6]. To simulate such a window, a population of heterogeneous synaptic compartments was modeled. Each synaptic compartment represents a single spine, one of the population of spines that corresponds to a stimulated empirical synapse. For this qualitative simulation, each of 40 synaptic compartments was coupled to its own dendritic compartment, and the dendritic compartments were independent of each other. Each compartment pair was identical to that used for the black curve in Fig. 5A2 except that a single parameter varied between pairs. The rate constant $k_{d \rightarrow s}$ for movement of PKM into the spine was varied (Fig. 5A2 legend). Spines with a higher $k_{d \rightarrow s}$ undergo LTP more easily. For these spines, $PKM_s$ switches to the upper state for greater absolute values of the WTET – STET interval (*i.e.*, less overlap between $T_{LTP}$ and $PKM_d$ time courses). The green curve in Fig. 5A2 shows the resulting LTP window. The average change in W over the population of spines is plotted. The window exhibits a sloped rise and fall, with nearly all the spines undergoing LTP for WTET – STET intervals near zero.

For the LTD window, a population of spines is not required, because in the model LTD is not based on a bistable switch. The amplitude is dependent on the degree of interaction of the variables $T_{LTD}$ and PRP, and the window has a graded rise and fall. In Fig. 5B2, significant LTD (which was defined as a decrease in W of 20% or more) occurs if WLFS precedes, or follows, SLFS by 75 min or less. This window width appears compatible with data in which LTD of 30 − 50% is seen when WLFS precedes SLFS by 40 min ([4], their Fig. 2H), and LTD between 22 − 40% when WLFS precedes SLFS by 60 min, with little LTD remaining for intervals of 2 or 3 h.

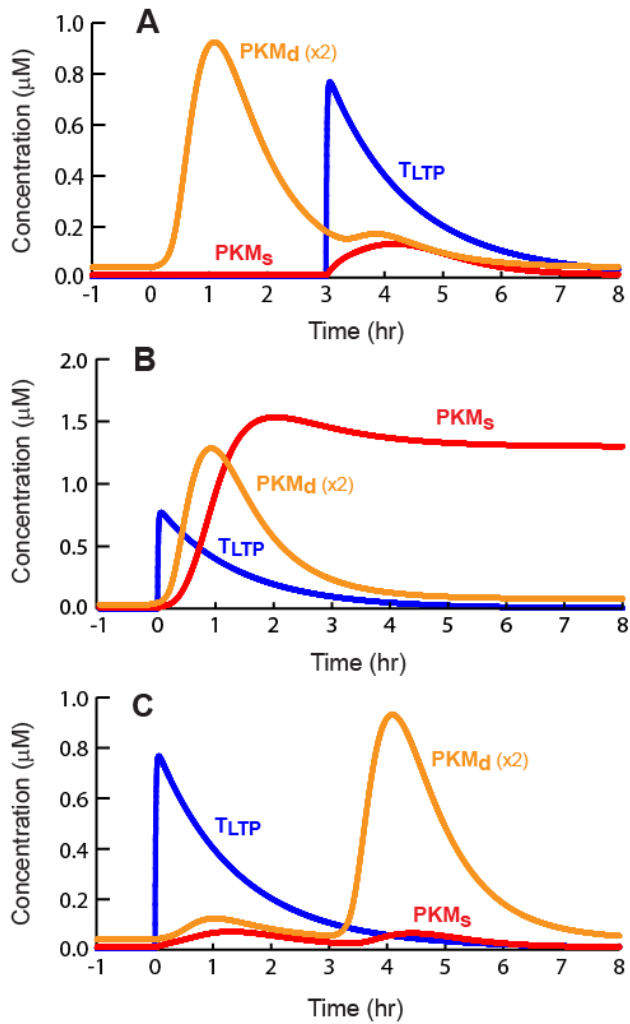

FIGURE 6  Simulated dynamics of key molecules in an STC protocol that pairs WTET and STET. *A,* Time courses of dendritic and synaptic PKM and the LTP tag, for WTET following STET by 3 h. *B,* WTET follows STET by 10 min. *C,* WTET precedes STET by 3 h.

Figure 6 illustrates the dynamics and overlap of elevations in PKM levels and the LTP tag, in WTET – STET STC. If STET to S2 precedes WTET to S1 by 3 h (Fig. 6A) or if STET follows WTET by 3 h (Fig. 6C), little overlap occurs in the elevations of $PKM_d$ with $T_{LTP}$. The product of $PKM_d$ and $T_{LTP}$ remains small, and by Eq. 29, little influx of $PKM_d$ into the synaptic compartment occurs. Only small transient elevations of $PKM_s$ result. However, if WTET and STET occur closer together (Fig. 6B), substantial overlap of $PKM_d$ with $T_{LTP}$ occurs, driving influx of $PKM_d$ into S1. $PKM_s$ increases sufficiently to initiate positive feedback.

The model also simulates late LTP and STC induced by chemical stimuli such as forskolin or BDNF [3,12], assuming forskolin and BDNF activate ERK and generate relatively low elevations of synaptic and dendritic $Ca^{2+}$. To simulate chem-LTP, $[Ca^{2+}]_s$ and $[Ca^{2+}]_d$ were increased to 0.24 μM for 30 min. For $k_{p\_Raf,s}$ and $k_{p\_Raf,d}$, maximal amplitudes were 0.007 min$^{-1}$ Both rate constants increased from basal to these amplitudes with a time constant of 0.5 min at the start of the 30-min stimulus, and decayed to basal with a time constant of 4 min after the end of the stimulus. Stable LTP resulted (a 169% increase in W 5 h post-stimulus). Subsequent inhibition of PKM by 80% for 1 hr reversed this LTP.

**Simulation of cross capture predicts LTD induction leads to dendritic synthesis of PKM**

Cross capture was simulated as WTET to S1 followed 20 min later by SLFS to S2. LTP of S1 resulted. Figure 7A illustrates time courses of $T_{LTP}$, $PKM_s$, $PKM_d$, and W. Setting the tag alone, by WTET, induces very little PKM synthesis. Therefore, the model assumes that SLFS must be present for the induction of dendritic PKM synthesis. Only then can PKM be captured at S1. The reverse experiment was also simulated, with WLFS to S1 followed 5 min later by STET to S2. Figure 7B illustrates that LTD is induced at S1. Only the LTD tag is set, capturing PRP but not PKM.

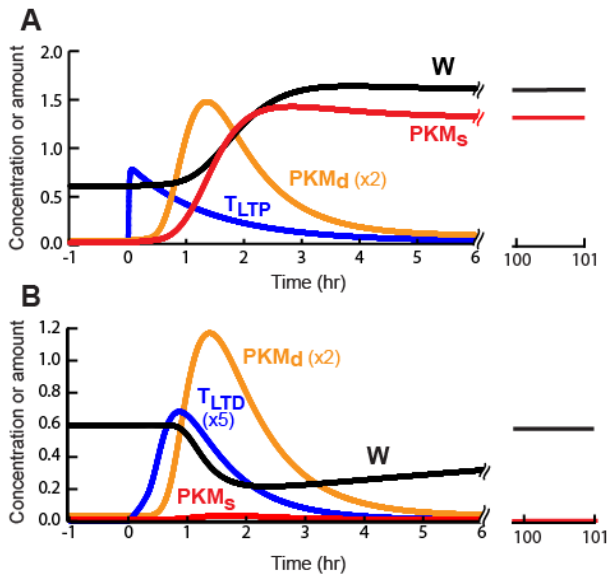

FIGURE 7  Simulation of cross capture. Vertical scaling factors are indicated next to the appropriate variable. The Δt for each simulation is 20 min. *A*, A WTET is delivered to S1 before SLFS is delivered to S2. Illustrated synaptic variables are for S1. The break in the x axis, followed by the traces for $PKM_s$ and W between 100 and 101 h, illustrates the permanent transition of $PKM_s$ and W to an upper state. *B*, A WLFS is delivered to S1 followed by a STET to S2. LTD results, but begins to decay after several h, as N and therefore W increase back towards baseline with the slow time constant $\tau_N$.

When the molecular identities of $T_{LTD}$ and $T_{LTP}$ are better characterized, it should be possible to test the model's prediction that following tetanic stimuli, $T_{LTD}$ remains low because activation of $ERK_s$ is insufficient to significantly phosphorylate an ERK tag site. The model predicts that during SLFS, $T_{LTD}$ increases over a time of minutes. In contrast, $T_{LTP}$ is set rapidly by a single tetanus (Fig. 2A). Because of this rapid setting, the model predicts that in a (WTET, STET) STC protocol, maximum LTP at S1 should occur when STET precedes WTET (Fig. 5A2, green curve). STET induces accumulation of PRPs, which are then available when WTET sets the tag.

**Stochastic simulations suggest positive feedback can sustain bistability of PKM activity in a realistic spine volume**

$PKM_s$ represents PKM dynamics in a synaptic compartment corresponding to a spine or a group of co-stimulated spines. Spines have small volumes, ~0.02 – 0.6 μm$^3$. Within each spine, concentrations of 1 μM would correspond to molecule copy numbers ranging from < 20 to a few hundred. For such numbers, random fluctuations in molecule numbers sometimes destabilize steady states, eliminating bistability [*e.g.* 21,22,23]. To examine whether the stable states of $PKM_s$ are robust to such fluctuations, stochastic simulations were performed (see Methods). Initially, the synaptic volume was set to 0.2 μm$^3$. To examine whether $PKM_s$ activity could be sustained solely by positive feedback, influx from the dendrite was removed by setting $T_{LTP} = 0$. Then $PKM_s$ dynamics are completely described by stochastic simulation of only Eq. 29, with the term containing $T_{LTP}$ removed and parameters at standard values.

Figure 8A illustrates that bistability was preserved. In each of 20 simulations, both the lower and upper steady states were stable for at least 3 days (black time course). The standard deviation (SD) of the 20 trajectories was not large. In the upper state, the average molecule numbers correspond to concentrations near 1.3 μM, similar to the upper state of the deterministic model (Fig. 2D, upper steady state of black curve). The basin of attraction for the upper state is substantial, extending well below the average molecule number of ~150. For $PKM_s$ initialized at 70, 18 out of 20 simulations converged to the upper state. However, for $PKM_s$ initialized at 35, 20 of 20 simulations fell to the lower state.

Figure 8

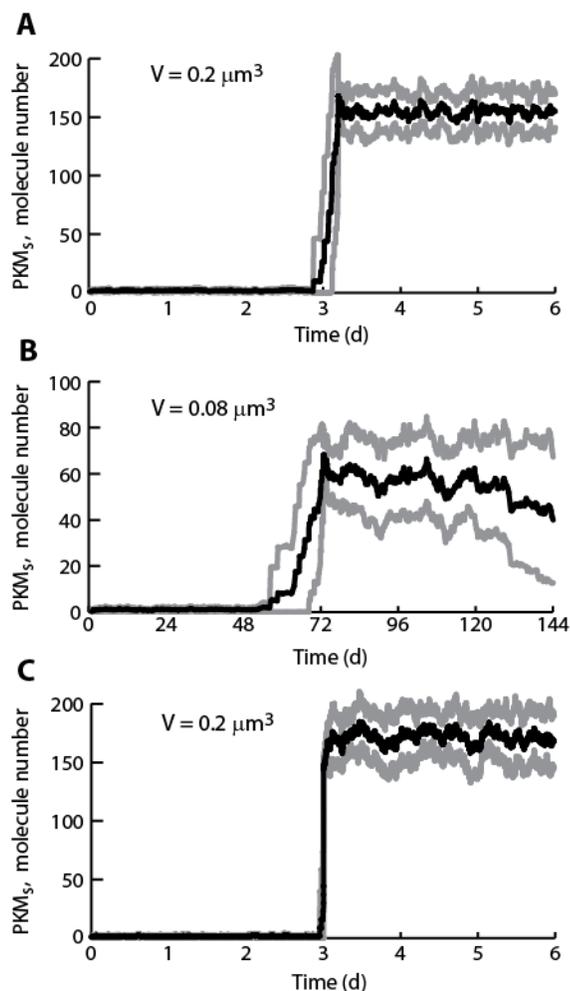

FIGURE 8  Stochastic simulations. **A,** For a volume of 0.2 μm³, both steady states are stable for at least 3 days. The black time course is for the average of $PKM_s$ over 20 simulations, in each of which $PKM_s$ was re-initialized to the upper steady state after the first ~72 h (the time of re-initialization varies somewhat between simulations because of variable time steps in the Gillespie algorithm). For all simulations the upper state remained stable for the next 3 days. The standard deviation of the trajectories over the 20 simulations was not large (grey time courses, ±1 SD from the average). **B,** For 0.08 μm³, the lower state remains stable, but the stability of the upper state degrades over several days. $PKM_s$ was re-initialized to the upper steady state after the first ~70 h. Over the following 3 days, the standard deviation increases as individual simulations fall back to the lower state. **C,** Similar to **A**, except that the rate of $PKM_s$ synthesis due to positive feedback is described by a series of elementary steps with first- or second-order kinetics instead of by a Hill function. Both steady states are again stable for at least 3 days.

If the volume was reduced to 0.08 μm³, the upper state was no longer stable for days. Figure 8B illustrates that for 20 simulations, the lower state was always stable for at least 3 days. However, following resetting, 5 of the simulations fell to the lower state during days 4-6. Thus the model suggests it is plausible that persistent activation of PKM can occur in larger spines (> 0.1 μm³, [24]).

Additional simulations were carried out to support the existence and robustness of bistability. If the simulation of Fig. 8A was repeated with a very large volume (200 μm³, $f_{stoch}$ = 120,000), fluctuations became very small. For both the lower and upper states, the mean concentrations of $PKM_s$ ±1 SD remained within 2% of the concentrations obtained in the corresponding deterministic simulation of Eq. 29. In addition, a model variant was constructed in which synthesis of $PKM_s$ induced by positive feedback is not described by a single Hill rate expression, but instead by a series of elementary unimolecular or bimolecular steps in which two PKM molecules bind sequentially to a target, with PKM synthesis only occurring after the second binding event (Methods). The off and on binding rates were chosen to yield a dissociation constant equal to the Hill constant in Eq. 29. Simulation of the elementary-step mechanism yielded lower and upper steady states of $PKM_s$ at molecule numbers similar to those of Figs. 8A-B. For V = 0.2 μm³ both steady states were again stable for at least 3 days (Fig. 8C).

## DISCUSSION

The model simulates dynamic elements of long-term synaptic plasticity including: 1) Nonlinear stimulus-response relationships for the activation of CaMKII and ERK. These nonlinearities generate large dynamic ranges of $CaMKII_s$, $ERK_s$ and $ERK_d$ activities in response to brief increases in $Ca^{2+}$ and Raf activation, thereby enabling brief stimuli to be transduced into long-lasting changes in synaptic weight, and 2) Convergence of multiple kinases to induce synthesis of PRPs (the variables PRP and $PKM_d$), and convergence of PRPs and synaptic tagging to induce LTP or LTD. This convergence ensures stimulus strength must exceed a threshold to induce LTP/LTD.

In neurons, there are likely to be numerous PRPs, some of which may contribute to either LTP or LTD depending on the identity of the synaptic tag. However, the current model illustrates that important aspects of STC, LTP, and LTD can be simulated using only two PRPs, one (PKM) specific for LTP and the second (PRP) specific for LTD.

**Key qualitative constraints appear necessary to simulate STC and cross capture**

To successfully simulate LTP/D, STC, and cross capture, specific model constraints were required, which have not yet been experimentally tested. These constraints can be summarized as follows. 1) In order for LTP to result from both STC and cross capture protocols, STET and SLFS must both induce synthesis of PKM in the dendritic compartment. However, PKM is only captured if an LTP tag is set. 2) Maintenance of LTP is mediated by bistability in PKM activity, with persistent PKM activation sustained by positive feedback restricted to the synaptic compartment, in which PKM enhances its synthesis. 3) A tetanus activates synaptic Raf and ERK, much less than dendritic Raf and ERK. This assumption prevents a tetanus from setting the synaptic LTD tag and inducing LTD. 4) A dendritic CaM kinase is required to induce PKM synthesis and is activated by moderate elevations of $Ca^{2+}$, such as are induced by SLFS. 5) A population of heterogeneous LTP compartments is required to simulate a graded LTP window (Fig. 5A2). This summary of constraints can serve as a reference for empirical investigations to test and delineate molecular mechanisms underlying STC.

Data appear consistent with the assumptions regarding PKM dynamics. Empirically, pairing WTET with SLFS induces LTP, but WTET alone does not induce LTP [4,5], suggesting PKM is synthesized in dendrites in response to either SLFS or STET, but is only captured at synapses tagged for LTP. Because of molecular turnover, a positive feedback loop appears necessary to create a bistable switch to sustain persistent PKM activity that maintains LTP. Restriction of bistability to the synaptic compartment is motivated by two considerations: 1) Empirically, PKM appears to accumulate in puncta at spines following LTP induction [19], rather than evenly in a dendrite, and 2) Dendritic bistability, with consequent persistently high dendritic PKM, would eliminate specificity in LTP expression, because all spines on a dendritic region would eventually undergo LTP in response to weak stimuli that set LTP tags. Persistently increased, punctate $PKM_s$ would require protein synthesis to occur in spines. Indeed, polyribosomes are found in spines [25] and their numbers increase after LTP induction [26]. In simulations including fluctuations in molecule numbers (Fig. 8), persistent activation of $PKM_s$ was maintained for at least several days in volumes typical of large spines on pyramidal neurons.

Limitations of previous models of STC have been delineated [3]. Our model appears to overcome some of these. First, tag setting and initial expression of LTP are separate and dissociable. If weak stimuli set tags but fail to synthesize PRPs, plasticity will not occur. Therefore, although tagging is necessary for expression of LTP or LTD, expression is not necessary for tagging. Second, determinants of both LTP and LTD are modeled. Only the LTP tag allows capture of PKM, whereas the LTD tag captures PRP only. Empirically, it is not known whether maintenance of LTD depends on persistent activity of an enzyme, analogous to PKM. Third, the model implicitly assumes an essential role for dendritic mRNA translation in the synthesis of $PKM_d$ and PRP in response to STET or SLFS, which activate dendritic ERK and CaM kinase. Several pathways exist for induction of translation by these kinases [27,28]. Fourth, the model simulates slow-onset synaptic potentiation induced by chemical stimuli such as forskolin or BDNF, by assuming these stimuli activate ERK and elevate synaptic and dendritic $Ca^{2+}$, thereby setting $T_{LTP}$ and activating PKM. Imaging of dendrites has demonstrated $Ca^{2+}$ elevations during BDNF application [29] and during intracellular cAMP elevation [30]. Recently, PKM activity was found necessary to maintain forskolin-induced LTP [31], as in the model.

The model does not represent all processes needed for tagging, such as cytoskeletal rearrangements [32], protein kinase A (PKA) anchoring [33]; and possible creation of slots for PRPs [34]. LTP also depends on PKA activation [35], and PKM synthesis is regulated by several kinases [11]. Thus, our representations of tagging and PRP synthesis are highly simplified. These simplified representations were chosen for a qualitative model that simulates the nonlinear stimulus-response relationships describing LTP/LTD induction, and LTP maintenance depending on persistent PKM activity. We also believe it is important that the model qualitatively represents the necessity for two compartments to describe PKM dynamics. To our knowledge, no other model has illustrated the need for a dendritic compartment in which PKM is synthesized, coupled with capture of PKM into a synaptic compartment and consequent persistent activity.

The model does not simulate metaplasticity in which a sliding threshold separating LTP induction from LTD induction depends on the history of postsynaptic activity, as in Bienenstock-Cooper-Munro theory [36]. Allowing parameters that govern the synthesis of LTP and LTD tags, and the synthesis of PRPs, to themselves depend on the activity history might allow simulation of such a threshold. Also, because the model does not represent sub-second kinetics of NMDA receptors, such as relief of $Mg^{2+}$ block, it cannot currently simulate spike timing-dependent plasticity.

**LTP maintenance is likely to depend on multiple, reinforcing feedback loops**

Auto-induction of PKM synthesis is unlikely to constitute the only positive feedback loop necessary to maintain LTP. Maintenance at late times also appears to rely on continued reactivation of NMDA receptors due to ongoing neuronal activity [37]. It is plausible such reactivation resets LTP and LTD tags, allowing for continued capture of PKM and other PRPs [38]. Interaction between CAMKII and NMDA receptors is needed to enhance spine growth [39] and maintain LTP [40]. Because activation of CAMKII induces its translocation to synaptic sites [41], it is plausible that continued reactivation of NMDA receptors is necessary to maintain this CAMKII – NMDAR interaction, and therefore LTP. Data from *Aplysia* suggest that aggregation of active cytoplasmic polyadenylation element binding protein may also be important for late maintenance of synaptic strength [42].

Fear conditioning memory is maintained despite inhibition of PKM in the amygdala [43] and the hippocampus [44]. Therefore, feedback loops other than those involving PKM may suffice to maintain some forms of memory. However, maintenance of spatial, instrumental, and classically conditioned memories is disrupted by inhibitors of PKM$\zeta$ [17] as are drug reward memory and avoidance responses [45,46].

Models suggest MAPK activity may exhibit bistability mediated by positive feedback [47]. LTP maintenance may also, in part, rely on feedback wherein transient activation of PKA phosphorylates and renders functional a critical number of AMPARs, sufficient to saturate phosphatases, so that basal PKA can subsequently maintain AMPAR phosphorylation [48]. Persistent activation of PKM together with another feedback loop with a shared common output (increase of functional AMPARs) could add robustness to LTP maintenance, generating a stable steady state that is more robust to stimulus fluctuations [49] and more robust to fluctuations in molecule numbers [50] and to inhibition of protein synthesis [51]. Further investigation of how feedback loops implicated in LTP interact to promote robustness and to maintain reliable induction, while filtering out fluctuations due to spontaneous activity, is important to understand how memory is induced and maintained.

**Molecular and phenomenological models of STC each have advantages**

Our model can be contrasted with two recent models [52,53] that also simulate aspects of STC. Both those models simulate STC and cross capture. They use variables that are more abstract or phenomenological than the molecular species of our model. In [53], Markovian transitions occur among synaptic states that differ in the presence or absence of tags for LTP/LTD and in the state lifetimes. A population of synapses onto one neuron is modeled in [52], which also represents LTP/LTD tags with bimodal variables, PRPs with a single variable p, and long-term maintenance of LTP with a state transition in a bistable variable z, the molecular nature of which is not specified. A strength of these more phenomenological models is that they can concisely represent and make clear the qualitative dynamics that must emerge from any plausible molecular schema for STC. We believe that our approach is complementary to the approach of those models, and has the strength of making predictions for the dynamics of specific molecules. Both modeling approaches should be pursued in tandem.

**The model suggests experiments to delineate mechanisms governing PKM$\zeta$ synthesis and LTP**

The model assumes SLFS induces PKM synthesis, in order to simulate LTP upon pairing of WTET and SLFS. However, empirically, an LFS of intermediate strength (3 Hz for 5 min) decreases PKM in hippocampal slice [54,55]. That intermediate LFS is weaker than the standard SLFS, and therefore may not elevate dendritic [Ca$^{2+}$] and ERK activity sufficiently to induce PKM synthesis. Nevertheless, consideration of these observations suggests a plausible alternate model. SLFS may only induce PKM synthesis if paired with WTET. In this scenario, SLFS would upregulate the activity of unspecified translation factors. The synaptic tag induced by WTET would include a propensity for increased synthesis of PKM$_s$, perhaps due to modified PKM mRNA structure or interactions. Increased synthesis of PKM$_d$ and increased PKM$_s$ would only occur if this propensity for increased PKM synthesis overlapped with upregulated activity of translation factors induced by SLFS. This alternate model is compatible with the induction of LTP by STC and cross capture protocols, whether WTET precedes or follows SLFS [4], as long as the tag overlaps with upregulation of translation factors.

A critical experiment, to distinguish between our model and this alternative, will be to examine whether PKM synthesis is induced by SLFS alone, or only by SLFS paired with WTET. Furthermore, if PKM synthesis can be induced by SLFS alone, its accumulation is predicted to be diffuse within a dendrite rather than punctate, due to the absence of tagged synapses that can capture PKM. In addition, if WTET is paired with SLFS, punctate accumulation should be observed.

The model assumes dendritic PKM synthesis depends on an unspecified CaM kinase, CK$_d$. This kinase is plausibly CaMKII, because application of KN-93, a CAMKII inhibitor, prevents the induction of PKM synthesis [11]. However, if SLFS alone is capable of inducing dendritic PKM synthesis, then CK$_d$ would apparently be activated at lower Ca$^{2+}$ levels than are required to activate CaMKII$_s$ (tetani have been shown to activate dendritic CaMKII [56,57], but SLFS has not). The experiments of [11] used 20 μM KN-93. At this concentration KN-93 may also inhibit other CaM kinases [58], thus this use of KN-93 has not firmly established CK$_d$ as CaMKII. These considerations suggest a further critical experiment, determination of whether a peptide inhibitor of CaMKII, such as autocamtide-2 related inhibitory peptide [59], blocks induction of PKM synthesis in a) a standard STET protocol, and in b) a cross capture protocol where WTET is paired with SLFS.

With our model, chem-LTP relies on a minor elevation of synaptic $Ca^{2+}$ elevation and CaMKII activation – enough to, over 30 min of stimulus, set the LTP tag. Therefore, CaMKII inhibition during stimulation with BDNF or forskolin is predicted to block LTP. BDNF or forskolin are also predicted to induce punctuate accumulation of PKM at spines.

Several mechanisms may underlie the positive feedback by which activity of PKM is maintained. PKM may catalyze enhanced translation of PKM mRNA, possibly via an intermediate step in which PKM represses Pin1 isomerase, thereby derepressing translation [60]. In another hypothesized feedback loop, synapses that have undergone LTP exhibit increased time-averaged synaptic activation [61] and increased average $Ca^{2+}$ levels [38], leading to increased activity of kinases that drive further PKM synthesis. Development of an expression construct that couples PKM mRNA elements with a fluorescent reporter, and examination of the reporter dynamics following electrical stimuli or glutamate applications, could help delineate the mechanisms of feedback.

## METHODS

Michaelis-Menten or first-order kinetics describe phosphorylations and dephosphorylations. Variables and parameters associated with the synaptic and dendritic compartments are respectively identified by the subscripts 's' and 'd'. The exceptions to this nomenclature are the LTP and LTD tags ($T_{LTP}$ and $T_{LTD}$) and W and its associated variables F and N, restricted to the synaptic compartment, and a representative plasticity-related protein, PRP, generated in the dendritic compartment. Units are μM for concentrations of molecular species, including phosphorylation sites. F, N, and W are dimensionless. The dynamics are described by 23 ordinary differential equations (ODEs) and auxiliary equations. Some of these equations are from our previous model of LTP induction [18]: Eqs. 1-17 describing activation of synaptic CaMKII, Raf, and ERK, and dendritic Raf and ERK; and Eq. 20 describing phosphorylation of the LTP tag site. Standard values for parameters are given in Methods, and were used in all simulations except where noted.

### Raf/MEK/ERK cascade

The ERK cascade is present in the dendritic ($Raf_d$/$MEK_d$/$ERK_d$) and synaptic ($Raf_s$/$MEK_s$/$ERK_s$) compartments (Eqs. 1-16). Total amounts of these enzymes are conserved (Eqs. 11-16). Stimuli elevate the rate constants for activation of $Raf_d$ and $Raf_s$, $k_{p\_Raf,d}$ and $k_{p\_Raf,s}$. The equations and parameters for the $ERK_d$ and $ERK_s$ cascades were identical, except that the two ERK cascades responded differently to stimuli, with different amplitudes of increase for $k_{p\_Raf,s}$ $vs.$ $k_{p\_Raf,d}$.

$$\frac{d}{dt}\left(pRaf_s\right) = k_{p\_Raf,s}\, Raf_s - k_{dp\_Raf}\, pRaf_s \tag{1}$$

$$\frac{d}{dt}\left(MEK_s\right) = -k_{p\_MEK}\, pRaf_s\, \frac{MEK_s}{MEK_s + K_{MEK}} + k_{dp\_MEK}\, \frac{pMEK_s}{pMEK_s + K_{MEK}} \tag{2}$$

$$\frac{d}{dt}\left(ppMEK_s\right) = k_{p\_MEK}\, pRaf_s\, \frac{pMEK_s}{pMEK_s + K_{MEK}} - k_{dp\_MEK}\, \frac{ppMEK_s}{ppMEK_s + K_{MEK}} \tag{3}$$

$$\frac{d}{dt}\left(ERK_s\right) = -k_{p\_ERK}\, ppMEK_s\, \frac{ERK_s}{ERK_s + K_{ERK}} + k_{dp\_ERK}\, \frac{pERK_s}{pERK_s + K_{ERK}} \tag{4}$$

$$\frac{d}{dt}\big(ppERK_s\big) = k_{p\_ERK}\, ppMEK_s\, \frac{pERK_s}{pERK_s + K_{ERK}} - k_{dp\_ERK}\, \frac{ppERK_s}{ppERK_s + K_{ERK}} \qquad (5)$$

$$\frac{d}{dt}\big(pRaf_d\big) = k_{p\_Raf,d}\, Raf_d - k_{dp\_Raf}\, pRaf_d \qquad (6)$$

$$\frac{d}{dt}\big(MEK_d\big) = -k_{p\_MEK}\, pRaf_d\, \frac{MEK_d}{MEK_d + K_{MEK}} + k_{dp\_MEK}\, \frac{pMEK_d}{pMEK_d + K_{MEK}} \qquad (7)$$

$$\frac{d}{dt}\big(ppMEK_d\big) = k_{p\_MEK}\, pRaf_d\, \frac{pMEK_d}{pMEK_d + K_{MEK}} - k_{dp\_MEK}\, \frac{ppMEK_d}{ppMEK_d + K_{MEK}} \qquad (8)$$

$$\frac{d}{dt}\big(ERK_d\big) = -k_{p\_ERK}\, ppMEK_d\, \frac{ERK_d}{ERK_d + K_{ERK}} + k_{dp\_ERK}\, \frac{pERK_d}{pERK_d + K_{ERK}} \qquad (9)$$

$$\frac{d}{dt}\big(ppERK_d\big) = k_{p\_ERK}\, ppMEK_d\, \frac{pERK_d}{pERK_d + K_{ERK}} - k_{dp\_ERK}\, \frac{ppERK_d}{ppERK_d + K_{ERK}} \qquad (10)$$

$$Raf_s + pRaf_s = TotRaf_s \qquad (11)$$

$$MEK_s + pMEK_s + ppMEK_s = TotMEK_s \qquad (12)$$

$$ERK_s + pERK_s + ppERK_s = TotERK_s \qquad (13)$$

$$Raf_d + pRaf_d = TotRaf_d \qquad (14)$$

$$MEK_d + pMEK_d + ppMEK_d = TotMEK_d \qquad (15)$$

$$ERK_d + pERK_d + ppERK_d = TotERK_d \qquad (16)$$

Standard parameter values for Eqs. 1-16 are as follows: $k_{p\_Raf,s}$ and $k_{p\_Raf,d} = 0.003$ min$^{-1}$ (basal value), $k_{dp\_Raf} = 0.12$ min$^{-1}$, $k_{p\_MEK} = 0.6$ min$^{-1}$, $k_{dp\_MEK} = 0.025$ μM min$^{-1}$, $K_{MEK} = 0.25$ μM, $k_{p\_ERK} = 0.52$ min$^{-1}$, $k_{dp\_ERK} = 0.025$ μM min$^{-1}$, $K_{ERK} = 0.25$ μM. $TotRaf_s$, $TotRaf_d$, $TotMEK_s$, $TotMEK_d$, $TotERK_s$, and $TotERK_d = 0.25$ μM.

For many parameters, standard values are not well constrained by current empirical data. Here and below, we note constraints that were used for some parameters. The remaining parameters were assigned standard values such that time courses of model variables had the qualitative properties needed to simulate STC and cross capture. In the model, standard total concentrations of Raf, MEK, and ERK are similar to estimates for neurons ([42], estimates range from 0.18 to 0.36 μM). Empirical estimates of the duration of ERK activation due to STET vary considerably (30 min – several hours [62,63]) such that the simulated duration of ~90 min is plausible.

ERK activity is not necessary for the LTP tag [64]. However, ERK activity is necessary for the induction of LTP [63,64,65]. ERK phosphorylates translation regulators including eukaryotic initiation factor 4E (eIF4E) [27]. Therefore the model assumes $ERK_d$ phosphorylates a site in the dendritic compartment that regulates the translation of PKM as well as of a second plasticity-related protein, PRP. ERK regulates some forms of hippocampal LTD [66,67] and is necessary for LTD tagging [64]. Thus in the model, translation of PRP is necessary for LTD, and $ERK_s$, in conjunction with PP, activates $T_{LTD}$.

## $Ca^{2+}$-dependent kinases

Two $Ca^{2+}$-dependent kinases are modeled: CaMKII in the synaptic compartment ($CaMKII_s$), and $CK_d$, a dendritic kinase that has yet to be characterized. CaMKII activity is necessary for LTP tagging [64]. The model assumes $CaMKII_s$ is responsible for phosphorylating a LTP tag ($T_{LTP}$). Tetani increase $Ca^{2+}$ sufficiently to activate $CaMKII_s$, but SLFS does not. Induction of PKM synthesis can be blocked by a calmodulin analogue, KN-93 [11]. Calmodulin analogues can bind to CaM kinases to form inactive complexes. However, we did not denote as CaMKII the dendritic CaM kinase required for PKM synthesis, because its identity is not yet firmly established (see Discussion). Therefore, an unspecified kinase $CK_d$, in conjunction with active $ERK_d$, induces the synthesis of PKM. For both $CaMKII_s$ and $CK_d$, $Ca^{2+}$-dependent activation is implemented as fourth powers in Hill functions.

$$\frac{d}{dt}\left(CaMKII_s\right) = k_{f\_CK,s}\ \frac{\left[Ca^{2+}\right]_s^4}{\left[Ca^{2+}\right]_s^4 + K1_s^4}\ - k_{b\_CK,s}\ CaMKII_s \tag{17}$$

$$\frac{d}{dt}\left(CK_d\right) = k_{f\_CK,d}\ \frac{\left[Ca^{2+}\right]_d^4}{\left[Ca^{2+}\right]_d^4 + K1_d^4}\ - k_{b\_CK,d}\ CK_d \tag{18}$$

Standard parameter values are: $[Ca^{2+}]_s$ and $[Ca^{2+}]_d = 0.04\ \mu M$ (basal value), $k_{f\_CK,s} = 200\ \mu M\ min^{-1}$, $k_{b\_CK,s} = 1.0\ min^{-1}$, $K1_s = 1.4\ \mu M$, $k_{f\_CK,d} = 200\ \mu M\ min^{-1}$, $k_{b\_CK,d} = 1.0\ min^{-1}$, $K1_d = 0.6\ \mu M$.

Empirically, binding of $Ca^{2+}$ to activate CaMKII is characterized by a dissociation constant of ~1-2 $\mu M$ [68], compatible with the standard value for $K1_s$. The decay of CaMKII activity after tetanus is rapid (~1 min [1]), compatible with the standard value of $k_{b\_CK,s}$. $CK_d$ has a lower dissociation constant ($K1_d$) for $Ca^{2+}$ than does CAMKII. The lower value of $K1_d$ was chosen so that SLFS, which elevates $Ca^{2+}$ less than STET, suffices to activate $CK_d$ and PKM synthesis.

## $Ca^{2+}$-dependent protein phosphatase

Empirically, inhibition of either PP2A [69] or PP2B [70] blocks hippocampal LTD. PRP synthesis is thought to subsume both LTP and LTD [3,7], whereas the type of tag ($T_{LTP}$ *vs.* $T_{LTD}$) confers specificity [64]. Because phosphatase inhibition has not been reported to block LTP, it is reasonable to assume such inhibition interferes with $T_{LTD}$. A phosphatase activity in the synaptic compartment, $PP_s$, is therefore assumed necessary to increase $T_{LTD}$. Stimuli increase $PP_s$ via increases in $Ca^{2+}$. $PP_s$ is a nonlinear function of $Ca^{2+}$.

$$\frac{d}{dt}\left(PP_s\right) = k_{f\_PP,s}\ \frac{\left[Ca^{2+}\right]_s^4}{\left[Ca^{2+}\right]_s^4 + K2_s^4}\ - k_{b\_PP,s}\ PP_s \tag{19}$$

Standard parameter values are: $k_{f\_PP,s} = 2$ μM min⁻¹, $k_{b\_PP,s} = 0.5$ min⁻¹, $K2_s = 0.225$ μM. The Ca²⁺ sensitivity of $PP_s$ ($K2_s$) is within the range of Ca²⁺ dissociation constant estimates for PP2B [71].

**LTP and LTD tags ($T_{LTP}$, $T_{LTD}$)**

$T_{LTP}$ is set by phosphorylation of a CaMKII$_s$ target site ("set" denotes increasing a tag variable from a low baseline to a substantial fraction of its maximal value). The amount of phosphorylated site is denoted $S_{CK}$, and is scaled to range from 0 to 1. The dynamics of $S_{CK}$ are described by a first-order ODE. The amount of $T_{LTP}$ is given by the square of $S_{CK}$. The square represents heuristically the likelihood that multiple phosphorylation events are necessary for tag setting.

$$\frac{d}{dt}\left(S_{CK}\right) = k_{p1}\,CaMKII_s\left(1 - S_{CK}\right) - k_{dp1}\,S_{CK} \tag{20}$$

$$T_{LTP} = \left(S_{CK}\right)^2 \tag{21}$$

$T_{LTD}$ is set by ERK$_s$ and PP$_s$. Setting $T_{LTD}$ requires phosphorylation of an ERK$_s$ target site and dephosphorylation of a separate site by PP$_s$. The ERK$_s$ site is denoted $S_{ERK}$ and the PP$_s$ site is denoted $S_{PP}$. $S_{PP}$ is increased by dephosphorylation. $T_{LTD}$ is proportional to the product of $S_{ERK}$ and $S_{PP}$.

$$\frac{d}{dt}\left(S_{ERK}\right) = k_{p2}\,ppERK_s\left(1 - S_{ERK}\right) - k_{dp2}\,S_{ERK} \tag{22}$$

$$\frac{d}{dt}\left(S_{PP}\right) = k_{dp3}\,PP_s\left(1 - S_{PP}\right) - k_{p3}\,S_{PP} \tag{23}$$

$$T_{LTD} = S_{ERK}\,S_{PP} \tag{24}$$

Rate constant values are chosen such that ERK$_s$ activates relatively slowly compared to CaMKII$_s$. This is necessary so that a brief tetanus does not set $T_{LTD}$ and generate LTD.

Standard parameter values for Eqs. 20-24 are: $k_{p1} = 0.45$ μM⁻¹ min⁻¹, $k_{dp1} = 0.006$ min⁻¹, $k_{p2} = 2.0$ μM⁻¹ min⁻¹, $k_{dp2} = 0.011$ min⁻¹, $k_{dp3} = 0.04$ min⁻¹, $k_{p3} = 0.011$ min⁻¹.

**Plasticity-related protein (PRP)**

Regulation and action of PRPs other than PKM were combined into a single variable, PRP. Induction of PRP synthesis was assumed to require phosphorylation by ppERK$_d$ at two dendritic sites. These sites are assumed to have the same kinetic parameters, so that a single variable, pTrans$_{ERK}$, describes either site. These two phosphorylations could represent activation of translation factors. The rate of PRP synthesis is taken as proportional to the square of pTrans$_{ERK}$. The square represents the requirement for phosphorylation of both sites. There is also a small basal rate of PRP synthesis, and first-order decay.

$$\frac{d}{dt}\left(pTrans_{ERK}\right) = k_{pERK}\,ppERK_d\left(1 - pTrans_{ERK}\right) - k_{dpERK}\,pTrans_{ERK} \tag{25}$$

$$\frac{d}{dt}\left(PRP\right) = k_{trans\_PRP}\left(pTrans_{ERK}\right)^2 + v_{bas\_PRP} - k_{d\_PRP}\,PRP \tag{26}$$

Standard parameter values for Eq. 26 are: $k_{pERK} = 4.0 \ \mu M^{-1} \ min^{-1}$, $k_{dpERK} = 0.1 \ min^{-1}$, $k_{trans\_PRP} = 2.2 \ \mu M \ min^{-1}$, $v_{bas\_trans} = 0.001 \ \mu M \ min^{-1}$, $k_{d\_PRP} = 0.022 \ min^{-1}$.

For Eqs. 20-26, parameter values are not constrained by data because the molecular identity of the LTP and LTD tags and of PRP(s) has not yet been characterized.

**Protein kinase M**

PKM dynamics are represented in synaptic and dendritic compartments by the variables $PKM_s$ and $PKM_d$. PKM mRNA is localized in dendrites [72], and inhibition of either ERK or CaM kinases prevents induction of PKM synthesis [11]. Therefore, in the model, synthesis of dendritic $PKM_d$ relies on activation of $ERK_d$ and $CK_d$. $ERK_d$ and $CK_d$ phosphorylate sites denoted by variables $pTrans_{ERK}$ and $pTrans_{CK}$. The dynamics of these sites are described by Eqs. 25 and 27.

$$\frac{d}{dt}\left(pTrans_{CK}\right) = k_{pCK} \ CK_d \left(1 - pTrans_{CK}\right) - k_{dpCK} \ pTrans_{CK} \tag{27}$$

Concurrent phosphorylation of these sites induces $PKM_d$ translation (Eq. 28). $PKM_d$ can also translocate to the synaptic compartment (Eqs. 28, 29) and constitutively active $PKM_s$ exerts positive feedback on its own synthesis (Eq. 29). The translocation rate is proportional to $T_{LTP}$, representing capture of PKM. Positive feedback leads to bistability and consequent persistently high $PKM_s$ levels. In Eq. 29, the feedback is nonlinear, with the synthesis rate proportional to a Hill function of $(PKM_s)^2$. Such nonlinearity is required for bistability, and could be generated if multiple $PKM_s$-mediated phosphorylations are required to induce $PKM_s$ synthesis. Bistability appears necessary to explain how a brief stimulus can give rise to persistent activation of PKM. In Eqs. 28-29, $V_{sd}$ denotes the volume ratio of the synaptic to dendritic compartments, with a standard value of 0.03.

$$\frac{d}{dt}\left(PKM_d\right) = k_{trans\_PKM,d} \ pTrans_{ERK} \ pTrans_{CK} - k_{d\rightarrow s} \ PKM_d \ T_{LTP}$$
$$+ k_{s\rightarrow d} \ V_{sd} \ PKM_s + v_{bas\_PKM,d} - k_{d\_PKM}PKM_d \tag{28}$$

$$\frac{d}{dt}\left(PKM_s\right) = k_{trans\_PKM,s} \frac{\left(PKM_s\right)^2}{\left(PKM_s\right)^2 + \left(K_{PKM}\right)^2} + \frac{k_{d\rightarrow s}}{V_{sd}} \ PKM_d \ T_{LTP}$$
$$- k_{s\rightarrow d} \ PKM_s + v_{bas\_PKM,s} - k_{d\_PKM}PKM_s \tag{29}$$

Standard parameter values for Eqs. 27 – 29 are: $k_{pCK} = 0.015 \ \mu M^{-1} \ min^{-1}$, $k_{dpCK} = 0.02 \ min^{-1}$, $k_{trans\_PKM,d} = 0.5 \ \mu M \ min^{-1}$, $k_{d\rightarrow s} = 0.0025 \ min^{-1}$, $k_{s\rightarrow d} = 0.012 \ min^{-1}$, $V_{sd} = 0.03$, $k_{d\_PKM} = 0.02 \ min^{-1}$, $k_{trans\_PKM,s} = 0.055 \ \mu M \ min^{-1}$, $K_{PKM} = 0.75 \ \mu M$, $v_{bas\_PKM,d}$ and $v_{bas\_PKM,s} = 0.0003 \ \mu M \ min^{-1}$. These parameters are not well constrained by data.

The model does not describe the dynamics of mRNA for PKM or PRP. Therefore it does not represent later, transcription-dependent phases of maintenance of LTP / LTD [73,74]. However, inhibition of transcription does not result in degradation of CA3-CA1 LTP until 5 - 8 h after induction [75,76]. The model does not simulate these times, but instead represents interactions of tags and PRPs during the first hours after stimuli. This time suffices to represent LTP formation and the establishment of persistently active $PKM_s$.

## Synaptic weight (W)

Changes in W represent empirical increases/decreases in excitatory postsynaptic potentials (EPSPs). Empirically, changes in W during induction of LTP or LTD, and during inhibition of PKM, occur on several time scales. Inhibition of PKM reverses LTP relatively rapidly, in ~ 1 h [15]. Late LTD can persist for several h or longer [20]. To implement these dynamics using a single synaptic weight variable would require a relatively small time constant for the return of W to baseline, in order to simulate the decay following PKM inhibition. However, the duration of LTD would then be limited to ~ 1 h by this time constant, unless LTD was maintained by its own bistable switch.

Instead of implementing a second bistable switch, an alternate approach was used. Eqs. 30-32 describe the dynamics of two variables, F and N, and set W as their product. Biophysically, this product could represent the number of functional AMPA receptors (AMPARs) in the synaptic spine (see below). F depends on $PKM_s$, and N depends on the LTD tag and therefore does not change significantly during LTP protocols. During LTP, F increases with $PKM_s$, and F, like $PKM_s$, relaxes to a stable upper state. N changes little, therefore W follows F. PKM activity is not necessary for LTD [16]. Thus, in the model, $PKM_s$ and F change little during LTD. However N, and therefore W, decreases when the LTD tag is set and PRP, the plasticity-related protein, is concurrently present. Eq. 30 gives the rate of decrease of N as proportional to the product of $T_{LTD}$ and PRP. This decrease of N and W is a simple representation of the empirical requirement for STC to drive late LTD [4,20]. There is also first-order decline in the number of AMPA receptors in the absence of replenishment.

$$\frac{d}{dt}(N) = -k_{LTD}\, T_{LTD}\, PRP(N) + v_{bas\_N} - \frac{N}{\tau_N} \qquad (30)$$

$$\frac{d}{dt}(F) = k_{LTP}\, PKM_s + v_{bas\_F} - \frac{F}{\tau_F} \qquad (31)$$

$$W = NF \qquad (32)$$

Standard parameter values are: $k_{LTD} = 0.03\ \mu M^{-2}\,min^{-1}$, $\tau_N = 600\ min$, $v_{bas\_N} = 0.0033\ min^{-1}$, $k_{LTP} = 0.014\ min^{-1}$, $\tau_F = 30\ min$, $v_{bas\_F} = 0.01\ min^{-1}$.

These standard values were chosen such that LTP induced by STET and LTD induced by SLFS both have magnitudes similar to empirical LTP and LTD [13], and develop over ~2 h, as does late LTP induced by BDNF/forskolin [12]. N was assigned a slow time constant so that following SLFS, LTD persists for several h. However, since N is not bistable, W eventually returns to its basal value. F was assigned a smaller time constant, to simulate observed kinetics of LTP decay after PKM inhibition.

Empirically, PKM maintains LTP by modifying trafficking of AMPARs to increase receptor insertion into postsynaptic sites [77,78]. During LTD, in contrast, dephosphorylation of Glu-R1 S845 appears to decrease the number of AMPARs underline available for insertion, by decreasing receptor abundance at extrasynaptic membrane or by destabilizing Glu-R1 homomers [79,80]. In the model, N represents the amount of available AMPARs. Assuming equilibration maintains a fixed ratio between the number of available receptors and inserted receptors, a simulated decrease in N (with F remaining approximately constant) represents a decrease in the number of inserted receptors and therefore represents LTD. In contrast, an increase in the variable F, driven by increased $PKM_s$ during LTP, represents an increase in the number of inserted receptors, without an overall change in the number of available receptors N. In all simulations, F remains below N. In Eqs. 30 and 31, small basal rates of increase of N and F are also present in order to sustain basal values of F, N, and W.

We believe this simplified model is reasonable given current data. However, it is undoubtedly incomplete. For example, LTP also correlates with phosphorylation of Glu-R1 S845 and Glu-R1 S831 [79], and the site(s) that PKM phosphorylates are not yet determined.

Although a recent model of LTP and LTD [52] assumed that a large number of spines (~50 or more) need to be activated to induce LTP, our basic model assumes the induction and expression of LTP and LTD occurs at single spines. Empirically, when glutamate was applied to a single spine, PRP synthesis and LTP were both observed [81]. We do expand the model to include a population of spines in simulations examining the extent of LTP as a function of the interval between S1 and S2 stimuli in STC.

**Compartment volumes and stochastic simulations**

The parameter $V_{sd}$ (Eqs. 28-29) denotes the ratio of the synaptic to dendritic compartment volumes. A standard value of 0.03 was chosen based on the following considerations: Radii of apical dendrites are typically ~0.25 μm [25,47,82]. Spine head volumes, which we identify with the synaptic compartment, vary widely, over a range of $0.02 - 0.6$ μm$^3$ [83]. In an STC protocol, dendritic diffusion of PKM and other plasticity-related proteins, near and between the two groups of spines corresponding to the stimulated inputs, would contribute to determining an effective $V_{sd}$. For protocols based on field stimuli, the typical distance scale between these groups of spines has not yet been determined. However, for an STC protocol based on stimulus of two separated single spines, LTP was observed to be near maximum for distances of $<\sim20$ μM [81]. If we adopt 20 μM as a relevant dendritic length scale, and use dendritic radii and spine volumes given above, a value of $V_{sd}$ in the range $0.01 - 0.1$ is suggested, with 0.03 being the midpoint on a log scale. Stochastic simulations suggest bistability of $PKM_s$, and stimulus-induced state transitions, can be preserved when $V_{sd}$ varies over this range, as long as the spine volume exceeds ~0.1 μm$^3$ (see Results).

Stochastic simulations used Eq. 29, with $T_{LTP} = 0$. For these simulations $PKM_s$ represents the number of active PKM molecules in a single spine. Fluctuations in $PKM_s$ were simulated using the Gillespie algorithm [84]. This algorithm takes variable time steps, and during each step exactly one reaction occurs. Which reaction occurs is determined randomly, with the probability of each reaction type proportional to its deterministic rate expression. In Eq. 29 with $T_{LTP} = 0$, there are four reaction types, with probabilities proportional to the four remaining terms on the RHS. These reactions are respectively synthesis of $PKM_s$ by positive feedback, efflux of $PKM_s$ into the dendrite, basal synthesis of $PKM_s$, and degradation of $PKM_s$. For the simulations of Figs. 8A-B, the Hill function in Eq. 29 was used directly to give the probability of $PKM_s$ synthesis instead of being replaced by elementary reaction steps. This simplification was motivated by a recent study [85] which found that for several models, direct use of Hill functions did not result in substantial differences in dynamics when compared to more complex simulations in which multiple elementary reactions were used. $PKM_s$ and the Hill constant $K_{pkm}$ were rescaled to molecule copy number by multiplication with a factor proportional to volume, denoted $f_{stoch}$. A spine volume of 0.2 μm$^3$ was assumed initially. In this volume 1 μM corresponds to 120 molecules, thus $f_{stoch} = 120$. The zero-order rate constants $k_{trans\_PKM,s}$ and $v_{bas\_PKM,s}$ were also multiplied by $f_{stoch}$. First-order rate constants were unchanged.

For the simulation of Fig. 8C, the Hill function was replaced by a series of elementary reaction steps, with parameters chosen to preserve similar dynamics to Fig. 8A. The Hill coefficient of 2 was replaced by a requirement for two PKM molecules to bind sequentially to an unspecified target species. Only when the target was fully occupied could PKM synthesis be induced. The total amount of target is denoted $T_{tot}$, and the amounts with one and two PKM molecules bound are denoted $T_1$ and $T_2$. In the Gillespie algorithm, the single reaction probability corresponding to positive feedback was replaced by the following five reaction probabilities:

Binding of PKM to the first target site, $k_{on} PKM_s (T_{tot} - T_1 - T_2)$

Dissociation of PKM from the first site, $k_{off} T_1$

Binding of PKM to the second site, $k_{on} PKM_s T_1$

Dissociation of PKM from the second site, $k_{off} T_2$

Synthesis of PKM, $k_{trans\_PKM,s} T_2$

Parameter values for these reactions were: $f_{stoch} = 120$, $T_{tot} = 1.5 f_{stoch}$, $k_{on} = 10.0 / f_{stoch}$, $k_{off} = K_{PKM} k_{on}$. This value of $T_{tot}$ yields a maximal $PKM_s$ synthesis rate 50% greater than that in the simulation of Fig. 9A. This rate increase was found necessary to maintain stability of the upper state.

**Simulating STC and cross capture**

Synaptic tagging and capture (STC) was simulated as follows. Synapse 1 (S1) is given a weak stimulus (WTET or WLFS). In the model, the dynamics of all synaptic variables take place at S1. S2 is given STET or SLFS. These STET or SLFS only affect dendritic variables, which primarily respond to the S2 stimulus, with a minor contribution from the weak stimulus at S1. In tagging or cross capture experiments, a time offset $\Delta t$, positive or negative, usually is placed between the S1 and S2 stimuli. For positive $\Delta t$, the S1 stimulus occurs first and sets either an LTP or LTD tag. Then, the S2 stimulus does not further affect the synaptic variables. However, PRP and $PKM_d$ synthesis are strongly induced. For either positive or negative $\Delta t$, only after both S1 and S2 stimuli will LTD occur at S1 (if $T_{LTD}$ was set by the S1 stimulus), or LTP occur (if $T_{LTP}$ was set).

**Simulation of stimulus protocols**

Stimuli elevate $[Ca^{2+}]$ and activate ERK signaling. Empirically, STET consists of three 1-s duration 100 Hz bursts of electrical stimuli, at intervals of 5 min. WTET is a single 1-s duration 100 Hz burst of activity. SLFS is a series of bursts, each of three stimuli at 20 Hz, with an inter-burst interval of 1 s, continuing for 15 min [6]. WLFS is a 15-min train of single stimuli at 1 Hz.

Details of $Ca^{2+}$ dynamics were not modeled. Instead, the $Ca^{2+}$ responses were modeled simply as square-wave increases from a basal level of 0.04 μM. Two $Ca^{2+}$ variables were used, the concentrations of synaptic and dendritic $Ca^{2+}$, $[Ca^{2+}]_s$ and $[Ca^{2+}]_d$. For STET, each 1-s tetanus was simulated as a 3-s increase of $[Ca^{2+}]_s$ to 1.4 μM and of $[Ca^{2+}]_d$ to 0.65 μM. This duration of $Ca^{2+}$ elevation appears compatible with data that imply a time constant in the range of 1-3 s for decay of $[Ca^{2+}]_s$ and $[Ca^{2+}]_d$ elevations after tetanus [86,87]. WTET is modeled as a single 3-s increase in $[Ca^{2+}]_s$ and $[Ca^{2+}]_d$ to these same values. SLFS was modeled as a 15-min elevation of $[Ca^{2+}]_s$ and $[Ca^{2+}]_d$ to 0.17 μM. WLFS was modeled as a 15-min elevation of $[Ca^{2+}]_s$ to 0.16 μM, and $[Ca^{2+}]_d$ remains at basal.

Each tetanus, SLFS, or WLFS generated transient increases in the first-order rate constants $k_{p\_Raf,s}$ and $k_{p\_Raf,d}$ that govern $Raf_s$ and $Raf_d$ phosphorylation and activation (Eqs. 1, 6). Increases in these rate constants correspond to increases in Ras activity. The neuronal Ras -> ERK cascade can be activated by $Ca^{2+}$ acting *via* CaM kinase I [88] or by cAMP elevation [89,90] or by a $Ca^{2+}$-independent pathway involving mGluR5 [91]. Rather than modeling these mechanisms in detail, we used data to approximate time constants for Ras activation and inactivation. Recent studies indicate the time required for Ras activity to increase to a peak following tetanus is ~ 1 min in dendrites and spines [92,93], although with considerable variability. Ras activity in both compartments then returns to baseline over the following 6-10 min, again with variability. These time courses were used as follows. For each tetanus, the simulated time courses of $k_{p\_Raf,s}$ and $k_{p\_Raf,d}$ were each represented as the product of two factors. The first was an exponential rise from a low basal value, $k_{p\_Raf,bas}$, to approach a maximal amplitude $A_{MAX}$, with a time constant of 0.5 min. The second factor was an exponential decay from 1 to 0 with a time constant of 4 min. For both exponentials, zero time was taken as the end of the 1-s tetanus. For the STET protocol, the decay time constant is similar to the interval between tetani, thus summation of the elevations of $k_{p\_Raf,s}$ and $k_{p\_Raf,d}$ was included. The equation for the time course of $k_{p\_Raf,s}$ due to STET is therefore

$$k_{p\_Raf,s} = k_{p\_Raf,bas} + \left(A_{MAX} - k_{p\_Raf,bas}\right)\sum_{i=1}^{3}\left\{1 - \exp\left(\frac{-t_i}{0.5}\right)\right\}\exp\left(\frac{-t_i}{4.0}\right) \tag{33}$$

In Eq. 33, the $t_i$ denote, respectively, the amounts of time since the end of tetani 1-3.

For LFS, the time course of Ras activation does not appear to have been quantified. Therefore, to model the elevations of $k_{p\_Raf,s}$ and $k_{p\_Raf,d}$ due to SLFS and WLFS, the same time constants were used as for tetanus. However, zero time for the rising exponential was taken as the beginning of the 15-min LFS, whereas the decaying exponential remained at 1 until the end of LFS. The time course of $k_{p\_Raf,s}$ due to SLFS or WLFS is therefore

$$k_{p\_Raf,s} = k_{p\_Raf,bas} + \left(A_{MAX} - k_{p\_Raf,bas}\right)\left\{1 - \exp\left(\frac{-t_1}{0.5}\right)\right\}\exp\left(\frac{-t_2}{4.0}\right) \tag{34}$$

In Eq. 34, $t_1$ and $t_2$ denote respectively the amounts of time since the beginning and end of an LFS. In Eqs. 33 and 34, each exponential was set to 1 whenever its $t_i$ was negative.

In spine and dendrite, $k_{p\_Raf,bas}$ was 0.003 $min^{-1}$. For tetani, the maximal amplitude $A_{MAX}$ of $k_{p\_Raf,s}$ was 0.006 $min^{-1}$. For SLFS and WLFS, $A_{MAX}$ of $k_{p\_Raf,s}$ was 0.02 $min^{-1}$. For $k_{p\_Raf,d}$, $A_{MAX}$ was 0.03 $min^{-1}$ (tetanus), 0.017 $min^{-1}$ (SLFS), or 0.006 $min^{-1}$ (WLFS). These amplitudes are not well constrained by current data. Therefore, the values were chosen not to fit data precisely, but rather to allow simulation of observed dynamics of STC and cross capture. For example, STET increases $k_{p\_Raf,d}$ more than $k_{p\_Raf,s}$, in order that substantial activation of ERK occurs in the dendrite but not in the spine. Thus PKM is synthesized in the dendrite, but the LTD tag is not set in the spine. With SLFS, $k_{p\_Raf,s}$ is lower than for STET. The lower value of $k_{p\_Raf,d}$ was chosen to avoid excessive activation of dendritic ERK by the longer SLFS stimulus, overproduction of PRP, and excessive LTD.

The assumption that STET activates synaptic Raf less than dendritic Raf could be valid if a specific configuration of Raf and other components of the ERK cascade is mediated through anchoring proteins. The configuration might generate slow kinetics at a step such as interaction of activated Ras with Raf. In that case, the synaptic ERK cascade might be activated only slightly by a brief tetanic stimulus, but strongly by the much longer SLFS.

## Numerical methods

For the simulations of Figs. 2 and 3, the ODEs were integrated by two methods, forward Euler and fourth-order Runge-Kutta [94]. No significant differences were observed in the results, therefore forward Euler was used for the remaining deterministic simulations. The time step was 12 ms. No significant improvements in accuracy resulted from further reducing the time step. Prior to any plasticity-inducing stimulus, model variables were equilibrated for at least two simulated days and the slowest variable, W, was set to an equilibrium basal value determined by the remaining variables. The models were programmed in Java. Programs are available upon request.

## ACKNOWLEDGEMENTS

We thank Harel Shouval for his comments on the manuscript. Supported by NIH grant R01 NS073974.